# Advanced Raman spectroscopy detection of oxidative damage in nucleic acid bases: probing chemical changes and intermolecular interactions in guanosine at ultralow concentration

Francesca Ripanti[1*], Claudia Fasolato[2*], Flavia Mazzarda[1†], Simonetta Palleschi[3], Marina Ceccarini[4], Chunchun Li[5], Margherita Bignami[3], Enrico Bodo[6], Steven E.J. Bell[5], Filomena Mazzei[3], Paolo Postorino[1]

[1] Department of Physics, Sapienza University of Rome, P.le A. Moro 5, Rome, Italy

[2] Department of Physics and Geology, University of Perugia, via Alessandro Pascoli, Perugia, Italy

[3] Department of Environment & Health, Istituto Superiore di Sanità, Viale Regina Elena 299, Rome, Italy

[4] National Centre for Rare Diseases, Istituto Superiore di Sanità, Viale Regina Elena 299, Rome, Italy

[5] School of Chemistry and Chemical Engineering, Queen's University of Belfast, Stranmillis Road, Belfast, Northern Ireland

[6] Department of Chemistry, Sapienza University of Rome, P.le A. Moro, 5, Rome, Italy

**ABSTRACT:** DNA/RNA synthesis precursors are especially vulnerable to damage induced by reactive oxygen species occurring following oxidative stress. Guanosine triphosphates are the prevalent oxidized nucleosites, which can be misincorporated during replication, leading to mutations and cell death. Here, we present a novel method based on micro-Raman spectroscopy, combined with *ab initio* calculations, for the identification, detection, and quantification of low concentrations of oxidized nucleotides. We also show that the Raman signature in the terahertz spectral range (< 100 cm$^{-1}$) contains information on the intermolecular assembly of guanine in tetrads, which allows to further boost the oxidative damage detection limit. Eventually, we provide evidence that similar analyses can be carried out on samples in very small volumes at very low concentrations by exploiting the high sensitivity of Surface Enhanced Raman Scattering combined with properly designed superhydrophobic substrates. These results pave the way for employing such advanced spectroscopic methods for quantitatively sensing the oxidative damage of nucleotides in the cell.

## INTRODUCTION

Reactive oxygen species (ROS), such as peroxides, superoxides, and hydroxyl radicals constitute a major source of damage to cellular components as lipids, proteins, and nucleic acids. They are the products of normal cellular metabolism, but also of exogenous stressors, such as ionizing radiation and chemicals. Damage occurs when an imbalance among ROS levels and cell antioxidant and repair capability is present, a circumstance that is generally termed oxidative stress [1-3]. ROS can interact with DNA and RNA molecules, resulting in modification of nitrogen bases [4,5], single- and double-breaks [6,7], abasic sites [8-10], and DNA/RNA-protein cross-links [11,12]. Among such ROS-induced harmful effects, DNA/RNA base oxidation is the most frequent damage.

More than 20 different types of oxidative damage to nitrogen bases have been identified [13,14]. Due to its low ionization potential [14,15], guanine is the base most susceptible to oxidation, commonly producing, in DNA and RNA, 8-oxo-7,8-dihydro-2'-(deoxy)guanosine (8-oxo-dG) and 8-oxo-7,8-dihydroguanosine (8-oxo-G), respectively. 8-oxo-dG (and related compounds) differs from the non-oxidized counterpart by the presence of a double C=O bond in position 8 of the aromatic ring, instead of a C-H bond. During DNA replication, 8-oxo-dG can pair with adenine instead of the canonical cytosine [5,15]. Whenever these lesions escape DNA repair processes, persistent 8-oxodG can result in GC→TA transversions [16]. Similar ROS-induced modifications are also found in 8-oxo-deoxyadenosine (8-oxo-dA) [17].

Several lines of evidence indicate that deoxyribonucleoside triphosphates (dNTPs) are relevant targets for oxidation in the nucleotide pool, mainly producing 8-oxo-7,8-dihydro-deoxyguanosine triphosphate (8-oxodGTP). Indeed, dNTPs are 13,000-fold more prone to oxidation than bases embedded in DNA [15] and the incorporation of oxidized dNTPs in the genome can result in mutagenic DNA damage [18]. Oxidation can also occur at the level of ribonucleoside triphosphates (NTPs), which are present in large excess over dNTPs in the nucleotide pool [19-21]. For example, since the concentration of guanosine triphosphate (GTP) is 100-fold higher than that of the deoxyguanosine triphosphate (dGTP) [22,23], significantly more 8-oxo-GTP than 8-oxo-dGTP is likely to be produced under oxidative stress conditions.

There is a great interest in identifying oxidized DNA and RNA lesions as biomarkers of oxidative stress, particularly in the case of isolated bases, whatever the origin of nucleic acid damage is, i.e. direct or mediated by the incorporation of modified precursors. Currently, the typical methods for detecting DNA/RNA base modifications include single-cell gel electro-phoresis assays [23-25], high performance liquid chromatography (HPLC) coupled with mass spectroscopy or electrochemical detection [26-28], and fluorescence staining techniques [29-31]. All these methods require multiple-step sample preparation and the use of chemicals that may induce additional modifications to the nitrogen base or interfere with the detection of the modified DNA/RNA. The need for robust, streamlined methods for tracing chemically modified DNA/RNA bases is an important goal for research, mainly in view of future clinical diagnostic applications and translational impact.

In order to address these limitations, here we propose an approach to detect and quantify the oxidized purine bases based on micro-Raman spectroscopy, with a particular focus on guanine. Raman spectroscopy is a non-invasive tool for biological diagnostics, providing direct information on the chemical composition and conformation of biomolecules without requiring the use of fixing agents or fluorescent probes. In the past few years, it has been extensively employed to characterize the chemical structure of DNA nucleotides [32-35], as well as to reveal changes in the structure of DNA, such as the formation of oxidative products [36-38]. Here, by combining Raman data with *ab initio* calculations, we demonstrate the quantitative detection of a low concentration of oxidatively damaged dNTPs and NTPs in standard solutions. We also show that the Raman signature of dGTP and GTP in the terahertz spectral range (< 100 cm$^{-1}$) contains information on their intermolecular assembly, which can be used in this context to further boost the oxidative damage detection limit.

Furthermore, we provide evidence that similar analyses can be conducted on samples in very small volumes and/or at very low concentrations, *i.e.* at conditions verified in cellular extracts, by exploiting the high sensitivity of Surface Enhanced Raman Scattering (SERS) [39-41], combined with properly designed superhydrophobic substrates [42,43]. SERS spectroscopy is based on the enhancement of Raman scattering signal from a specific analyte either adsorbed or placed in close proximity to a nanostructured noble metal surface. This occurs because the energy of the laser used for Raman excitation, normally in the visible spectral range, is in resonance with the collective excitation of the free electrons (localized surface plasmon resonance) in the metal nanostructures [39-41,44]. Light scattering by the metal nanostructure causes an amplification of the electromagnetic field in the proximity of the metal surface, which yields in turn a strong enhancement of the Raman signal (by factors up to $10^{10}$) from molecules nearby. The possibility of SERS detection of specific molecules at ultralow concentration is well established in the literature [40,45,46], and the coupling of SERS with superhydrophobic substrates, to efficiently exploit microvolume samples for spectroscopic analysis, has been successfully realized in the last decade [43,47,48]. Here, we report a novel SERS strategy for the quantitative detection of oxidative damage in sub-microliter nucleotide pool solutions, exploiting properly designed and functionalized superhydrophobic needles. Our method allows reliable measurements with detection limits suitable for the biologically relevant values [28]. We believe that these results represent a promising starting point for the development of advanced spectroscopic assays to detect the presence of oxidized nucleotides in cellular dNTPs/NTPs pools, possibly discriminating normal and cancer cells.

**EXPERIMENTAL SECTION**

**Materials**

dGTP, 8-oxo-dGTP, 8-oxo-GTP, and 8-oxo-dATP were purchased from Jena Biosciences GmbH (Jena, Germany), GTP was purchased from Promega (Madison, WI, United States), and dATP, deoxyguanosine (dG), and 8-oxo-dG from Sigma-Aldrich (St. Louis, MO, United States). All the nucleotides and nucleosides were in the stable sodic salt form.

**Mixture preparation**

Using 1 mM nucleotide standard solutions in DNAse-free water, 8-oxo-dGTP and 8-oxo-GTP was diluted at different relative concentration in dGTP and GTP solution respectively, ranging from 16% to 0%. Higher 8-oxo-dGTP molar fractions were not considered since such high values were not revealed in biological samples [28]. Following this method, mixtures of dATP and 8-oxo-dATP at different relative concentrations were also prepared. Since, to the best of our knowledge, no data are available on the quantification of oxidized adenosine in biological samples, mixtures were prepared at varying oxidized/non-oxidized nucleotide molar ratios in a larger range (50%-0%).

Measurable samples for Raman spectroscopy were obtained by drop-casting a microvolume (5 μl) of the aqueous solution under examination on a flat gold substrate. We stress that the starting concentration of the nucleotides plays a minor role considering that the Raman spectra were acquired on dried samples (15 minutes drying at ambient conditions). To obtain measurable samples for SERS experiments, the analyte was mixed with a solution of hydroxylamine reduced silver nanoparticles, a 1 μL droplet was deposited on the tip of the superhydrophobic wire and measured thereon.

**Superhydrophobic substrate fabrication**

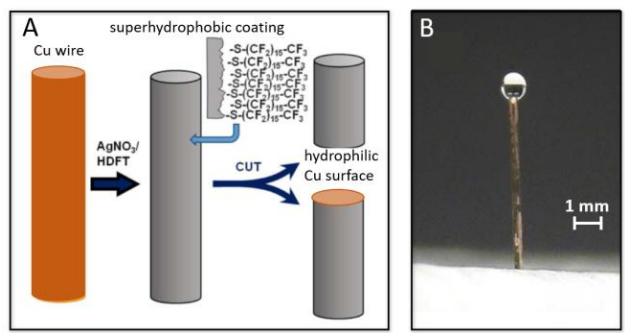

**Figure 1** (A) Schematic diagram of the experimental procedure for the fabrication of superhydrophobic substrates; (B) photograph of a superhydrophobic needle with a sample droplet on the tip. The volume of the droplet is 1 μL. Figure adapted from Ref. [43].

The superhydrophobic supports were realized using galvanic deposition on copper wires. The wires (400 μm diameter) were immersed in 0.01 M AgNO$_3$ aqueous solution for 1 minute, which gave a matt black textured silver surface coating, and dried. The metal-coated wire was then placed into a 0.01 M solution of a polyfluorinate surface modifier such as HDFT

(3,3,4,4,5,5,6,6,7,7,8,8,9,9,10,10,10-heptadecafluoro-1-decanethiol) in dichloromethane for 2 minute and dried. Then, the superhydrophobic coated wire was cut using a sharp scalpel to expose bare copper, which would act as the hydrophilic tip and hold aqueous samples [42,43]. As droplet dispenser we used a gas chromatography syringe, whose needle was given a superhydrophobic coating, so that the dispensed volume readily detached from the needle when it was placed in contact with the tip of the wire support. The syringe needle was coated by first electrodepositing a copper layer at 1.5 V in a simple cell containing $CuSO_4$ (1 M) acidified with $H_2SO_4$ and a clean copper foil counter electrode. The copper surface of the syringe was then coated with electroless deposited silver and polyfluorothiol, following the same protocol used for covering the copper wire. This coated syringe allows sub-µL volume liquid samples to be easily transferred on the support tip. When the droplet was brought into contact with the hydrophilic region of the superhydrophobic wire and the needle, the force holding the droplet to the syringe was less than that holding it to the support tip. A schematic representation of the fabricated superhydrophobic copper needles is shown in Figure 1.

### Silver nanoparticle preparation

The hydroxylamine reduced silver colloid was prepared using a well-established protocol [40,49]. Briefly, 5 mL of NaOH (0.1 M) was added to 5 mL of aqueous hydroxylamine hydrochloride (6 mM), then the whole mixture added to 90 mL of aqueous $AgNO_3$ (0.1 mM) with stirring. The colloid formed spontaneously and was left stirring for about 20 minutes before use.

### Micro-Raman spectroscopy

Raman measurements were carried out using a Horiba HR-Evolution micro-spectrometer in backscattering geometry, equipped with a He-Ne laser, λ=632.8 nm and 25 mW output power (≈10 mW at the sample surface). The elastically scattered light was removed by a state-of-the-art optical filtering device based on three BragGrate notch filters [50], which also allows to collect Raman spectra at very low frequencies (down to 10 $cm^{-1}$ from the laser line). The detector was a Peltier-cooled charge-coupled device (CCD) and the resolution was better than 3 $cm^{-1}$ thanks to a 600 grooves/mm grating with 800 mm focal length. The spectrometer was coupled with a confocal microscope supplied with a set of interchangeable objectives with long working distances and different magnifications (100x - 0.80 NA was used for the present experiment). Further details on the experimental apparatus can be found in Ref. [51]. All the spectra were pre-processed with LabSpec software (polynomial baseline subtraction in the fingerprint region, linear background in the low frequency range) and analyzed with Origin Lab code.

### SERS spectroscopy

Measurements on needles were performed using the Perkin Elmer RamanMicro 200. This apparatus consists of a 785 nm external cavity diode laser, outputting 90 mW via fiber optic cable to an Olympus BX51 Reflected Illumination microscope, which is equipped with 10x, 20x, and 40x objective lens with different numerical aperture and laser spot size. The samples were supported on a manually operated standard microscope stage. The scattered light was collected at 180° through the objective lens and was passed down a separate collection fiber towards the spectrograph, which was based on a Czerny-Turner design. The CCD detector was an electrically cooled Andor DV 420 OE and operated at -50°C. The instrument had a fixed resolution of 8 $cm^{-1}$ and spectra were collected over the 115-3200 $cm^{-1}$ range. For these experiments 10x objective (0.25 N.A. and 0.1 mm spot diameter) was used and laser power was set to 30% (higher power would burn the samples). Accumulation times were usually varied depending on the sample.

Diluted dG and 8-oxo-dG were mixed with the nanoparticle solution in 1:10 volume ratio, using $MgSO_4$ as aggregating agent. The final concentrations of the mixtures were in the $10^{-4}$-$10^{-7}$ M range. SERS spectra were analyzed using the partial least squares (PLS) regression through the PLS-1 package of GRAMS software. A standard approach was applied: the spectra were pre-processed by taking Savitsky-Golay first derivative (13 points, $2^{nd}$ degree) to suppress the effect of variations in the background. The spectral range between 400-1700 $cm^{-1}$ was chosen since it contains the main bands of dG and 8-oxo-dG (and relative compounds), excluding the characteristic peak of the colloid (around 240 $cm^{-1}$, data not shown) and spectral artefacts. Different types of intensity normalization were explored, and the normalization by the whole spectral area was selected. Three components were sufficient to reconstruct the experimental data by PLS regression.

### *Ab initio* calculations

Calculations were performed in the gas phase using the neutral structure corresponding to the bisodic salt of the monoprotonated triphosphate anion for both the dGTP and the 8-oxo-dGTP. Given the large conformational flexibility of such molecules, the conformational space was explored by means of molecular dynamics (MD). In particular, 3 short independent MD trajectories of 5 ps were recorded using the DFTB+ package [52] with SCC charges [53], dispersion interactions [54], and the 3ob parameter set. From each trajectory, a set of 3 structures were extracted for a total of 18 structures. These structures were then optimized at the B3LYP/6-31G* level using the g16 package [55]. We then selected the 7 optimized structures with lowest energy (all structures turned out to lie within 10 kcal/mol with respect to the lowest energy one) for which we computed harmonic frequencies and Raman activities. The harmonic frequencies have been scaled by 0.985 to account for anharmonicity (two of these structures are reported in Figure S1 in Supporting Information).

## RESULTS

### Raman spectra of purine deoxy-nucleoside triphosphates
### Analysis of 8-oxo-dGTP/dGTP

A combined theoretical and experimental analysis allowed to precisely characterize the different vibrational modes of the oxidized and non-oxidized nucleotides. Indeed, the different molecular structures of 8-oxo-dGTP and dGTP (Figure 2) result in different intramolecular electronic distributions, yielding a significantly different Raman response in the fingerprint region (900-1700 $cm^{-1}$), where the spectral bands associated to vibrations within the molecule are present.

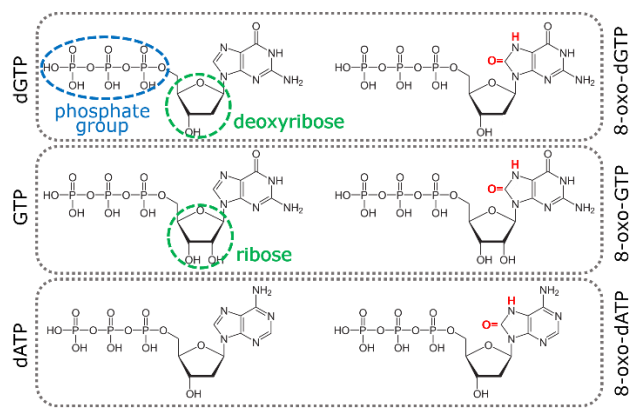

**Figure 2** Molecular structures of the analyzed nucleoside triphosphates (left) and their oxidized counterparts (right). From the top: deoxyguanosine triphosphate, guanosine triphosphate, and deoxyadenosine triphosphate. The different sugar component is highlighted in green, while the oxidative damage is marked in red.

To identify the vibrational modes associated to the presence of the extra oxygen in 8-oxo-dGTP, Density Functional Theory (DFT) calculations were carried out. A comparison between calculated and measured spectra for both the dGTP and 8-oxo-dGTP pure samples is reported in Figure 3. The theoretical calculations are consistent with the literature and allowed us to assign the main spectral bands.

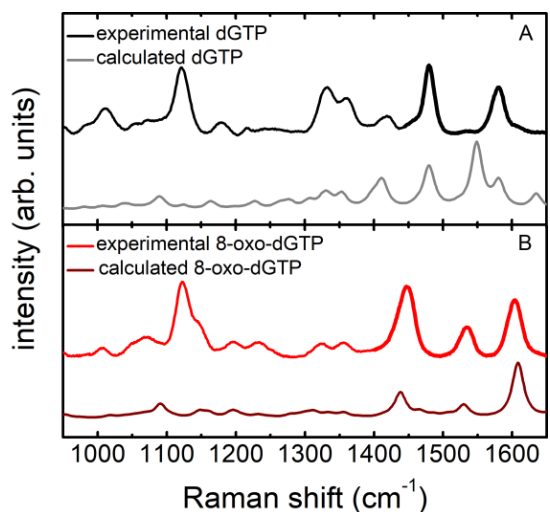

**Figure 3** Comparison of experimental and calculated Raman spectra: (A) dGTP and (B) 8-oxo-dGTP. The spectral regions exploited for the quantitative analysis are marked in bold.

Theoretical and experimental data show a remarkable agreement, despite some differences in the frequency values and relative intensities are present, mainly in the case of the dGTP sample. It is worth to notice that in the experimental spectra a strong band at $\nu_{PO} = 1123$ cm$^{-1}$ is clearly detectable. According to our theoretical calculations and literature data [40], this feature can be ascribed to the PO$_4^{3-}$ stretching vibration of the nucleotide phosphate backbone. Thus, the peak can be used for spectra normalization.

The main effect of oxidation is easily recognizable in the 1400-1650 cm$^{-1}$ spectral range (Figure 3). The 8-oxo-dGTP spectrum (panel B) shows three well-defined and separated bands, whereas in the dGTP spectrum (panel A) only two peaks are present. According to our DFT calculations on dGTP, the band centered at 1485 cm$^{-1}$ is ascribed to the 7N-8C stretching vibration and the one at 1575 cm$^{-1}$ to the 3N-4C stretching mode. In 8-oxo-dGTP, instead, the band at 1445 cm$^{-1}$ is associated to the 7N-H bending vibration, the one at 1535 cm$^{-1}$ to the 7N-8C stretching mode of the aromatic ring, and the third peak at 1607 cm$^{-1}$ is ascribed to the 5C-7N stretching vibration. These data are in good agreement with previous theoretical [56-58] and experimental [37,38,59] work.

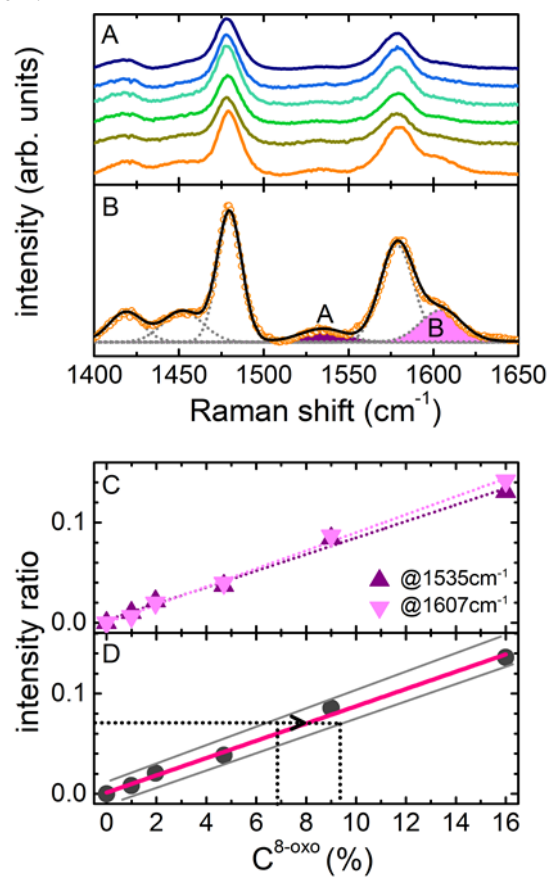

**Figure 4** Raman analysis of nucleotide mixtures in the fingerprint spectral region: (A) Raman spectra of a mixture of oxidized and non-oxidized dGTP solution at different 8-oxo-dGTP/dGTP relative concentrations, from 0% (blue, top) to 16% (orange, bottom); (B) representative fitting deconvolution at $C^{8-oxo} = 16\%$: hatched areas identify the peaks of 8-oxo-dGTP used for the quantitative analysis; (C) normalized integrated peak intensity (indicated as intensity ratio) as a function of 8-oxo-dGTP concentration for A (1535 cm$^{-1}$) and B (1607 cm$^{-1}$) peaks (purple and magenta, respectively), and linear fit of the data as described in text, with $m_A = 0.008 \pm 0.001$ and $m_B = 0.009 \pm 0.001$; (D) calibration curve obtained as the average of the two peak intensities. The 8-oxo-dGTP/dGTP relative concentration is inferred from the measured intensity $I_i$ (see text for details). The associated uncertainty was calculated from the intercept with the calibration and error curves.

A careful characterization of the spectroscopic response of oxidized and non-oxidized nucleotides through Raman spectroscopy allows, in principle, for a quantitative study of the oxidation level in cellular nucleotide pools. We thus analyzed artificial oxidized/non-oxidized mixtures at controlled relative concentrations (the 0-16% range was explored based on biological considerations [28]). Spectra are shown in Figure 4A. As we assumed the same Raman cross-section for dGTP and 8-oxo-dGTP, we normalized the data to the intensity of the triphosphate band ($v_{PO}$ = 1123 cm$^{-1}$). A deconvolution of the spectra with six Gaussian curves was carried out in the 1400-1650 cm$^{-1}$ spectral region. The agreement between fitting curve and the experimental data is excellent for any of the measured mixtures, as shown for a representative sample ($C^{8-oxo}$ = 16%) in Figure 4B. In the normalized spectra, the intensity of the two 8-oxo-dGTP bands, centred at $v_A$ = 1535 cm$^{-1}$ and $v_B$ = 1607 cm$^{-1}$ respectively (see the colored curves in Figure 4B), measures the oxidized base content in the mixture. The Raman intensity vs relative concentration trend, derived from fitting, is described by the linear function

$$I_i = C_i^0 + m \cdot C_i^{8-oxo}$$

where i=A,B identifies the vibrational peak with intensity $I_i$, $C^{8-oxo}$ is the 8-oxo-dGTP relative concentration in the mixture, and $C^0$ is the intercept at $C^{8-oxo}$ = 0, *i.e.* the value obtained from the fitting procedure for peaks A and B on the dGTP spectrum. By a close inspection of Figure 4A, a weak spectral contribution at the characteristic 8-oxo-dGTP frequencies can be observed in the spectrum of the dGTP alone: this might be ascribed to the accidental presence of a $C^0$ concentration of 8-oxo-dGTP in the dGTP sample. The $I_i$ data retrieved from fitting of the mixture spectra, subtracted by the contribution corresponding to $C^0$, are shown in Figure 4C along with the $I_i$ vs $C_i^{8-oxo}$ calibration curves for A and B bands (intercept = 0 by construction). The calibration lines for A and B are parallel, with the same slope within uncertainties. This confirms that the analysis can be safely considered as a spectroscopic measurement of the relative 8-oxo-dGTP/dGTP content. The average calibration curve (from A and B) is shown in Figure 4D, with nominal error estimated as the deviation of the experimental data from the average values. Again, by construction, the error is zero for the dGTP alone ($C^{8-oxo}$ = 0). The calibration curve is easily used for assessing the unknown oxidized nucleotide content in a mixture, as shown schematically in Figure 4D. The proposed protocol yields a detection limit of the 1% in terms of relative 8-oxo-dGTP/dGTP concentration.

It is worth noting that in the last few years many efforts have been devoted to the identification of a procedure for the detection and quantification of oxidized nucleotides. Pursell and coworkers [28] used HPLC with electrochemical detection to assess the 8-oxo-dGTP content in mitochondrial extracts from rat liver, heart, brain, skeletal muscle, and kidney, and compared those data with measurements of the four canonical dNTPs in the same extracts. They demonstrated that the 8-oxo-dGTP/dGTP ratio is tissue-dependent and ranges from 0.7% and 10% [28]. Therefore, the sensitivity of our Raman-based approach is comparable to their lower limit.

**Analysis of 8-oxo-dATP/dATP**

In order to verify whether a similar method is also suitable for detecting other damaged nucleotides, we characterized the spectroscopic response of 8-oxo-2'-deoxyadenosine-5'-triphosphate (8-oxo-dATP) and repeated the same analysis on oxidized/non-oxidized adenosine mixtures (see Supporting Information). Similar to 8-oxo-dGTP, 8-oxo-dATP differs from its non-oxidized counterpart (dATP) by the presence of an oxygen atom in position 8 of the aromatic ring and of a hydrogen atom bound to the nitrogen in the position 7 (Figure 2) [17,60]. The oxidized nucleotides show a significantly different Raman spectrum with respect to the non-oxidized ones, although a well detectable and isolated peak ascribed to the oxidation is not easily identified. The comparison of dATP and 8-oxo-dATP Raman spectra in the fingerprint region is shown in Figure S2A. Based on our analysis, the best oxidation marker is the spectral band around 620 cm$^{-1}$, ascribed to the C5−N7−C8 squeezing vibration [61,62], which is present almost exclusively in the 8-oxo-dATP spectrum. Following the same procedure as above, mixtures of 8-oxo-dATP/dATP at different relative concentrations were prepared and the corresponding Raman spectra collected. In this case, we explored a larger 8-oxo-dATP/dATP relative concentration range with respect to the 8-oxo-dGTP case. The spectral weight of the 620 cm$^{-1}$ band was evaluated through an accurate fitting procedure and the normalized integrated areas (subtracted by the value relative to the dATP alone) as a function of 8-oxo-dATP percentage are reported in Figure S2B. A clear linear dependence on concentration was observed and data analysis in the fingerprint region provides a sensitivity of oxidized nucleotide detection lower than 2%. Although this value is comparable with that obtained for 8-oxo-dGTP, the 8-oxo-dATP experimental data are more dispersed from the linear fitting curve. This could be explained considering the spectral range of the selected 8-oxo-dATP marker. In the case of 8-oxo-dATP, a single isolated peak was not available for the analysis and a dATP-associated spectral background produces some noise in the quantitative measurements. Nevertheless, the protocol yields results that are consistent with the case of guanine compounds, thus allowing for the quantification of the oxidation level also for adenine samples.

**Raman spectra of ribonucleoside triphosphates**
**Analysis of GTP/8-oxo-GTP**

Since our general aim is to propose sensitive methods for the quantification of oxidized nucleotides in cellular pools, which are dominated by the presence of ribonucleotides, we carried out our analysis also on ribonucleotide mixtures, with different relative concentrations of GTP and 8-oxo-GTP. NTPs are characterized by the presence, instead of deoxyribose, of a ribose sugar unit, which differs for the presence of an oxygen atom bound to the 2-carbon (Figure 2). The type of the sugar does not affect the spectroscopic response, at least in the fingerprint spectral range, as suggested in previous work [63]. The Raman spectra of the single GTP/8-oxo-GTP and dGTP/8-oxo-dGTP nucleotides look very similar (Figure S3). This result suggests that the quantitative analysis can be extended to the ribonucleotide mixtures.

The spectra of GTP and 8-oxo-GTP samples reproduced the significant differences observed on the deoxy counterpart with the same spectral markers of oxidation, namely the two peaks centered at $v_A$ = 1535 cm$^{-1}$ and $v_B$ = 1607 cm$^{-1}$ (Figure 5A). 8-oxo-GTP/GTP mixtures at different relative concentrations (0-16% range, as in the deoxy case) were spectroscopically analyzed. The same procedure (normalization, subtraction of pure GTP contribution, and linear fitting) yielded the calibration curve reported in Figure 5B, with a slope value in agreement with that obtained for the 8-oxo-dGTP/dGTP mixtures. This demonstrates that the Raman signature of oxidation is not affected by the presence of different sugars. For both cases (dGTP and GTP), the oxidative damage detection limit is assessed around the 1% oxidized/non-oxidized relative concentration.

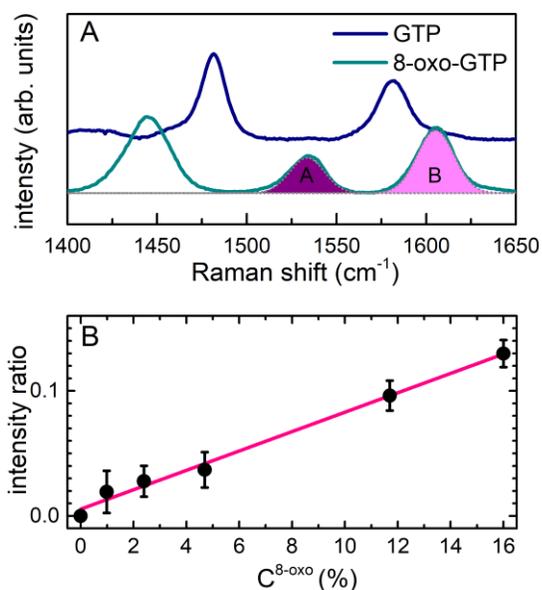

**Figure 5** (A) Comparison of Raman spectra of GTP (blue, top) and 8-oxo-GTP (dark cyan, bottom) in the 1400-1650 cm$^{-1}$ spectral region. The spectral markers of oxidation, A and B bands at 1535 and 1607 cm$^{-1}$, are fitted with Gaussian curves and evidenced by the colored areas; (B) calibration curve obtained as the average of two integrated intensities (A, B) vs $C^{8-oxo}$. Error bars are estimated by the partial dispersion of repeated measurements. The slope is $m_{ribo} = 0.008 \pm 0.001$.

**Monitoring intermolecular organization by terahertz Raman spectroscopy**

At high concentration, guanine compounds are known to form self-ordered tetrad aggregates, usually referred as G-tetrads or G-quartets, which tend to stack in columnar structures [64-66]. These structures are made of piles of planar tetramers, stabilized by eight Hoogsteen and Watson–Crick hydrogen bonds (N1–O6, N2–N7), with the O6 atoms oriented centrally into the ring to form an anionic bipyramidal cage that can coordinate to monovalent cations (e.g. Na$^+$, K$^+$). G-tetrads are present also in G-rich DNA sequences, in crucial domains of the genome, and in RNA sequences [67-70], due to the folding of DNA/RNA strands. These structures are termed G-quadruplexes. The presence of oxidative damage can interfere with the formation of such intermolecular organization, hindering its biological role. Indeed, the different structure and electronic distribution in oxidized nucleotides produce structural rearrangements at the intermolecular level, modifying the extent and the properties of long-range intermolecular interactions, which are reflected in the terahertz Raman response. In the past years, the characterization of the self-assembled structures has been carried out exploiting nuclear magnetic resonance spectroscopy, scanning electron microscopy, dynamic light scattering, and Raman in the fingerprint region [63, 71-73]. To the best of our knowledge, the low frequency Raman response of intermolecular G structures has not been reported yet and has the advantage of containing intense and easily monitored spectral markers of tetrads formation.

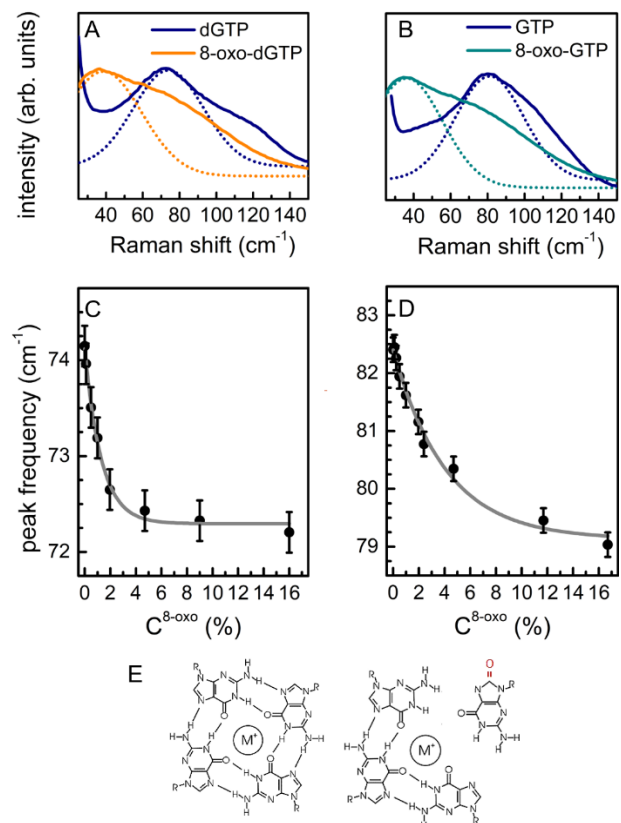

**Figure 6** (A) Comparison of the THz Raman spectra of dGTP and 8-oxo-dGTP, along with the main component of the gaussian fitting deconvolution; (B) same for GTP and 8-oxo-GTP; (C) frequency of the peak shown in (A) in 8-oxo-dGTP/dGTP mixtures as a function of the $C^{8-oxo}$ relative concentration; (D) same for 8-oxo-GTP/GTP. (E) On the left, a schematic representation of a G-tetrad. On the right, sketch of the hindered tetrad formation due to the presence of 8-oxo-G: we speculate that some of the H bonds constructing the tetrad cannot form, preventing its construction.

Exploiting the remarkable performance of Volume Bragg filters for ultralow frequency detection of the Raman signal [50], we developed a complementary spectroscopic strategy for oxidative damage quantification, by focusing on the low-frequency (THz) Raman response. By collecting the Raman spectra in the very low frequency region, i.e. down to 20 cm$^{-1}$, the spectral features associated with intermolecular organization can be probed. The THz Raman spectra (20-200 cm$^{-1}$) of dGTP and 8-oxo-dGTP are shown in Figure 6A. Both spectra are characterized by a broad spectral structure, but they differ by a remarkable (> 30 cm$^{-1}$) shift in the frequency of the main spectral band. The same occurs in the spectra of GTP and 8-oxo-GTP (Figure 6B). The fitting deconvolution of the spectra are shown in Figure S4. A gradual frequency shift depending on oxidized/non-oxidized relative concentration is observed in the mixture spectra, as shown in Figures 6C and 6D for deoxy-and ribonucleotides, respectively. Remarkably, in both cases, the main peak frequency $\nu_{C^{8-oxo}}$ as a function of $C^{8-oxo}$ is well described by an exponential curve:

$$\nu_{C^{8-oxo}} = \nu_0 - (\nu_0 - \nu_{8-oxo})\, e^{-R \cdot C^{8-oxo}}$$

where $\nu_0$ and $\nu_{8\text{-}oxo}$ are the asymptotic values for GTP and 8-oxo-GTP, respectively. The fitting provides two different values for the parameter R in the two cases (deoxy- and ribonucleotides), which imply two different decay rates for the exponential curves. This effect can be reasonably attributed to the role of the sugar (deoxyribose and ribose) in the long-range intermolecular interactions [63].

We hypothesize that this exponential behavior is associated to the tetrad formation process. Low-frequency vibrational modes, considered as collective intermolecular modes, are expected in the THz spectral region. The low-frequency features observed on the dGTP/GTP samples (Figures 6A and 6B) can thus be attributed to vibrations associated to the G tetrad structure [70]. A different THz Raman feature is observed in the oxidized samples. Due to the presence of an O8 atom, oxidized guanine can form different bonds with the neighbouring N6, which is involved in the tetrad H-bonds, hence impeding the aggregation process, and thus tetrad structures cannot form in these solutions (sketch in Figure 6D). The presence of supramolecular aggregates in this case is still under study and continuous helical structures have been proposed [74]. This hypothesis explains the large difference between the oxidized and non-oxidized molecule spectra in the low-frequency region. In the case of mixtures, the presence of oxidized molecules interferes with the tetrad formation. Indeed, at very low oxidation levels, molecules are mostly ordered in aggregates, while, at increasing oxidation level, some of the tetrads are expected to break down or their formation is going to be inhibited, thus decreasing the number of aggregates. Remarkably, the analysis of intermolecular effects provides a detection limit around the 0.2%, therefore boosting by an order of magnitude the sensitivity obtained from the fingerprint Raman signal analyses.

**Oxidized nucleoside detection in microvolume samples by Surface Enhanced Raman Scattering**

We showed that Raman spectroscopy allows discriminating low concentrations of oxidized dNTPs/NTPs in a non-oxidized pool, paving the way to applications on biological samples. However, with current purification procedures, the extraction of dNTPs/NTPs from the cellular cytoplasm provides very diluted samples in very small volumes. To enable the application of the proposed spectroscopic method to biological samples, we exploited the enhanced sensitivity of SERS spectroscopy. We also used superhydrophobic substrates to further reduce the minimum sample volume required for the analysis. Differently from the standard strategies for preparing synthetic or artificial superhydrophobic materials, including chemical reactions, chemical vapor deposition, nanolithography, electrospinning, layer-by-layer self-assembly, and phase-separations [75,76], we exploited the alternative method proposed in Refs. [42,43,47], which is based on a cost-effective, convenient, and time-saving approach. This method is indeed based on electroless galvanic deposition of silver from an aqueous solution onto copper metal. Details on the substrate preparation are provided in the Methods section. Following this approach, superhydrophobic copper needles were fabricated, as shown schematically in Figure 1, and then used as substrates for SERS measurements, which were performed on 1 μL droplets of solution.

As discussed in the Methods, the hydroxylamine-stabilized Ag colloid was prepared according to a well-established protocol [40,49], resulting in a negatively charged colloid. Since dGTP and 8-oxo-dGTP nucleotides (and their RNA counterpart) are negatively charged due to the presence of the phosphate groups, to favor a better electrostatic interaction between the sample and the colloid, the SERS measurements were performed directly on dG and 8-oxo-dG. As a preliminary step, we made sure that the Raman spectrum of the nucleosides preserved the characteristic spectral features of oxidation.

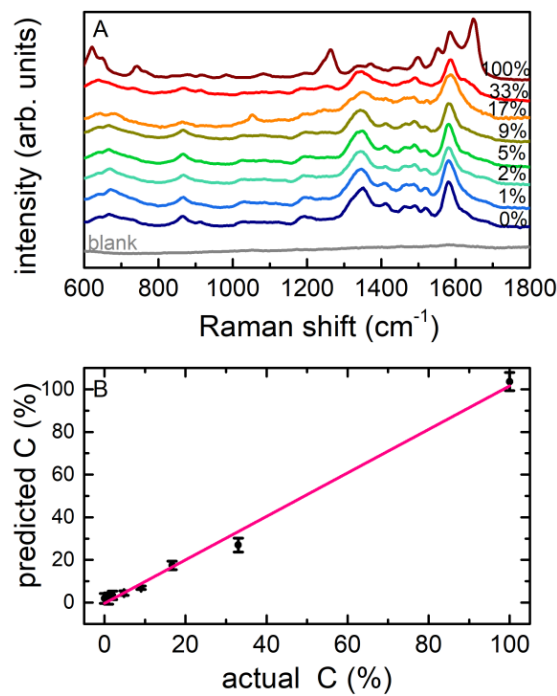

**Figure 7** (A) SERS spectra of nucleoside mixtures at varying $C^{8-oxo}$ from 0% (solid blue line, bottom) to 100% (wine, top) acquired on superhydrophobic substrates. Note that the spectra are not corrected by any baseline subtraction. The spectrum of the blank colloid solution is reported for comparison (dashed line); (B) multivariate data analysis of the SERS spectra: PLS regression plot of predicted vs actual $C^{8-oxo}$. The error bars show the standard deviation of 3 repeated measurements.

As a proof-of-principle experiment to verify the sensitivity and the reproducibility of the spectra acquired on the needles, decreasing concentration of dG were first analyzed (Figure S5). At least for concentrations down to $10^{-7}$ M, high-quality spectra were collected. It is notable that such analysis allows the measurement of extremely small amounts of molecules. Indeed, a 1 μL droplet at $10^{-7}$ M analyte concentration corresponds to sub-picomolar detection. Mixtures were prepared and analyzed following the procedure described above, and the obtained SERS spectra are shown in Figure 7A as a function of the relative content of 8-oxo-dG.

Some spectroscopic markers of the presence of 8-oxo-dG can be identified by visual inspection of the mixture spectra (e.g. the spectral features around 1250 cm$^{-1}$ and the region be-tween 1450 and 1700 cm$^{-1}$). However, their quantification is not trivial even by spectral deconvolution, because of the intrinsic complexity of SERS spectra [77-80]. Partial least squares (PLS) regression resulted as a more successful approach, allowing to discriminate the spectral features of the oxidized/non-oxidized molecules. PLS analysis is based on a model where a few spectra are employed to create a calibration curve that can be used in quantitative analysis (details provided in the Methods section). The model-predicted 8-oxo-dG concentrations are plotted against the actual values in Figure 7B. The data show a linear behavior and a sensitivity around the 1% in the 8-oxo-dG content is reached. These data are in complete agreement with the results of the fingerprint Raman analysis, but allow for a remarkable reduction of both the sample volume and the concentration. It is important to notice that, in this experiment, mixed samples were prepared at a final concentration of $10^{-4}$ M, but similar high-quality spectra can be obtained even at much lower concentrations of guanosine alone (Figure S5). These findings prove, in principle, the feasibility of 8-oxo-dG femtolmoles detection.

### Conclusions

In conclusion, we developed diverse Raman spectroscopy-based assays for the detection of oxidative damage in a nucleotide pool, mainly exploring the biologically relevant case of oxidized guanine. Compared to other diagnostic methods, Raman spectroscopy is realized by a single step measurement protocol, rapidly providing chemically specific signatures of the detected molecules. The discussed methods can be extended to other types of oxidative damage to DNA/RNA nucleotides, as demonstrated in the case of adenine.

When looking at the Raman spectra in the fingerprint spectral region, where bands associated to the vibration of specific chemical groups are found, some spectral markers of oxidation can be identified. DFT calculations coupled to Raman spectroscopy allowed the specific assignment of the vibrational modes associated to oxidation. By studying the fingerprint Raman spectral response of artificial nucleotide mixtures, we demonstrated that 8-oxo-G/8-oxo-dG can be revealed in a pool of non-oxidized G/dG down to relative concentrations as low as the 1%, which can be quantitatively determined by a thorough spectral deconvolution analysis.

Furthermore, terahertz Raman spectroscopy, which provides information on intermolecular organization, enabled monitoring the formation of tetrad G-aggregates. The characterization of the low frequency Raman signature associated to the formation of G-tetrads is, to the best of our knowledge, a novel result of the present work that might be exploited for studying G-quadruplex structures for a wide variety of biological applications. Our results on mixtures of oxidized/non-oxidized molecules suggest that the presence of damaged bases in the nucleotide pool hinders the formation of tetrads, resulting in a modified THz Raman response. This can be analyzed for quantitatively assessing the percentage of oxidized nucleotides well below the 1% detection limit.

To further enhance the sensitivity of our analysis, we turned to surface enhanced Raman spectroscopy, which we exploited in combination with superhydrophobic substrates to spectroscopically probe microliter sample volumes. We characterized the SERS signature of dGTP/GTP and 8-oxo-dGTP/8-oxo-GTP and demonstrated that SERS measurements provide high quality spectra from a μl droplets down to sub-μM concentration. The SERS detection limit of 8-oxo-dG in an oxidized/non-oxidized nucleoside pool is around 1%, which is well sufficient for biomedical applications. Based on these results, we speculate that the detection of oxidized guanosine femtomoles is feasible, paving the way to applications exploiting the ultrasensitive detection of oxidized guanine in cells as biomarkers for early events possibly leading to cancer.

### AUTHOR INFORMATION

**Corresponding Authors**

* Francesca Ripanti: francesca.ripanti@uniroma1.it
* Claudia Fasolato: claudia.fasolato@unipg.it

**Present Addresses**

†Frank Reidy Research Center for Bioelectrics, Old Dominion University, Norfolk, VA, USA

**Author Contributions**

All authors have given approval to the final version of the manuscript.

### ACKNOWLEDGMENTS

F. Ripanti acknowledges Sapienza University of Rome for financial support through the International Mobility Fellowship of her visiting period to Queen's University in Belfast.